\documentclass{svmult}

\usepackage{graphicx}    

\def\beq{\begin{equation}}
\def\eeq{\end{equation}}

\def\bea{\begin{eqnarray}}
\def\eea{\end{eqnarray}}
\def\ba{\begin{array}}                  
\def\ea{\end{array}}

\begin{document}

\title*{Seiberg-Witten maps and anomalies in noncommutative Yang-Mills theories}
\titlerunning{Seiberg-Witten maps and anomalies} 
\author{Friedemann Brandt}
\authorrunning{F. Brandt} 
\institute{Max-Planck-Institute for Mathematics in the Sciences,\\
Inselstra\ss e 22-26, D-04103 Leipzig, Germany\\
\texttt{}}
%
%
\maketitle
\begin{abstract}
A BRST-cohomological analysis of Seiberg-Witten maps and results on gauge anomalies in noncommutative Yang-Mills theories with general gauge groups are reviewed. 
\end{abstract}

\section{Introduction}
\label{intro}

We shall discuss two aspects of noncommutative Yang-Mills theories
of the type introduced in \cite{Jurco:2000ja} 
(see sect. \ref{rev} for a brief review). The first aspect concerns
the construction of these theories which is based on so-called 
Seiberg-Witten mappings (SW maps, for short). 
These mappings express `noncommutative'
fields and gauge transformations in terms of the standard
(`commutative') fields and gauge transformations. The mappings
have been named after Seiberg and Witten because they were
established first in \cite{Seiberg:1999vs} for the particular
case of $U(N)$-theories. However, it should be kept
in mind that in the present context they are
not limited to $U(N)$-theories but extended to
other gauge groups. This raises the questions whether and why
SW maps exist for general gauge groups, how they can be constructed 
efficiently and to which extend they are unique resp. ambiguous.
These questions are the topic of sect. \ref{sect3} which
reviews work in collaboration
with G. Barnich and M. Grigoriev 
\cite{Barnich:2002tz,Barnich:2002pb,Barnich:2003wq}.

Sect. \ref{sect4} reports on work in collaboration with
C.P. Mart\'{\i}n and F. Ruiz Ruiz
\cite{Brandt:2003fx}. It 
addresses the question whether the gauge symmetries
of noncommutative Yang-Mills theories can be anomalous when one applies the standard perturbative approach to (effective) quantum field theories.
It is not to be discussed here whether or not such an approach makes sense;
currently there is hardly an alternative perspective on these theories 
in the general case (i.e., for a general gauge 
group) since the theories are constructed only by means of SW maps and
no formulation in terms of `noncommutative' 
variables is known.
Hence, at present we have to content ourselves with a formulation of the `effective type' that
is not renormalizable by power counting, i.e., a
Lagrangian containing field monomials of arbitrarily high mass dimension.
As a consequence, there is no simple argument which can rule out 
from the outset the possible occurrence
of gauge anomalies with mass dimensions larger than 4. This complicates the
anomaly discussion as compared to renormalizable Yang-Mills theories 
whenever the gauge group
contains at least one abelian factor since in that case there is 
an infinite number of
candidate gauge anomalies
in addition to the well-known chiral gauge anomalies.


\section{Brief review of noncommutative Yang-Mills theories}\label{rev}

The noncommutative Yang-Mills theories under consideration involve
a $\star$-product given by the 
Weyl-Moyal product,
\begin{eqnarray*}
f_1 \star f_2 = f_1\, \exp(\stackrel{\leftarrow}{\partial_\mu}
\frac{{\mathrm{i}}}{2}\,\tau\,\theta^{\mu\nu}
\stackrel{\rightarrow}\partial_\nu)\,f_2,
\quad
\theta^{\mu\nu}=-\theta^{\nu\mu}=\mbox{constant}.
\end{eqnarray*}
$\tau$ is a constant deformation parameter that has been
introduced for the sake of convenience.
The `noncommutative' generalization of the Yang-Mills action
reads
\beq
\hat I[\hat A]=
-\frac 14\int d^nx\, \mathrm{Tr}\, (\hat F_{\mu\nu}\star\hat F^{\mu\nu}),\quad
\hat F_{\mu\nu}=\partial_\mu\hat A_\nu-\partial_\nu\hat A_\mu+
\hat A_\mu\star\hat A_\nu-\hat A_\nu\star\hat A_\mu
\label{r1}
\eeq
where $\hat A_\mu$ is constructed
from `commutative' gauge potentials $A_\mu$
by means of a SW map. $A_\mu$ lives in the Lie algebra
of the gauge group and has the standard Yang-Mills gauge transformations,
\beq
\delta_{\lambda}A_\mu=\partial_\mu\lambda+[A_\mu,\lambda]\equiv D_\mu\lambda,
\label{r2}
\eeq
where $\lambda$ denotes Lie algebra valued gauge parameters.
SW maps, by definition, express the noncommutative gauge potentials
$\hat A_\mu$ and gauge parameters $\hat \lambda$ in terms of $A_\mu$ and $\lambda$
such that (\ref{r2}) induces the noncommutative version of
Yang-Mills gauge transformations given by
\beq
\hat \delta_{\hat \lambda}\hat A_\mu=
\partial_\mu\hat \lambda+\hat A_\mu\star\hat \lambda-\hat \lambda\star\hat A_\mu\equiv\hat D_\mu\hat \lambda.
\label{r3}
\eeq
Furthermore we require that $\hat A_\mu$ and $\hat \lambda$ coincide with
$A_\mu$ and $\lambda$ at $\tau=0$,
\begin{eqnarray*}
\hat A_\mu=\hat A_\mu(A,\tau)=A_\mu+O(\tau),\quad
\hat \lambda=\hat \lambda(\lambda,A,\tau)=\lambda+O(\tau).
\end{eqnarray*}
Hence, SW maps are required to fulfill
\begin{eqnarray*}
\delta_{\lambda}\hat A_\mu(A,\tau)
=(\hat \delta_{\hat \lambda}\hat A_\mu)(A,\tau).
\end{eqnarray*}
For the inclusion of fermions see, e.g., \cite{Jurco:2000ja,Barnich:2002pb}.


\section{Analysis of SW maps}\label{sect3}

Noncommutative Yang-Mills
theories can be regarded as consistent deformations of corresponding
commutative Yang-Mills theories. This allows one to apply BRST-cohomological
tools to analyse SW maps along the lines of \cite{Barnich:2001mc}. In the following, we first review briefly
the BRST-cohomological approach to consistent deformations and then
the results on SW maps.

\subsection{Consistent deformations}

Consider an action $I^{(0)}[\varphi]$ with
gauge invariance
$\delta^{(0)}_\lambda$, i.e. $\delta^{(0)}_\lambda I^{(0)}[\varphi]=0$.
Consistent deformations of $I^{(0)}[\varphi]$ and $\delta^{(0)}_\lambda$
are power series' $I[\varphi,\tau]$ and $\delta_\lambda$ in
a deformation parameter $\tau$, such that
the deformed action  is invariant under the (possibly) deformed gauge transformation,
\begin{eqnarray*}
I[\varphi,\tau]=I^{(0)}[\varphi]+\sum_{k\geq 1}\tau^k I^{(k)}[\varphi],
\quad
\delta_\lambda=\delta^{(0)}_\lambda+\sum_{k\geq 1}\tau^k \delta^{(k)}_\lambda,
\quad
\delta_\lambda I[\varphi,\tau]=0.
\end{eqnarray*}
Two such deformations are called equivalent ($\sim$) if they are
related by mere
field redefinitions $\hat \varphi(\varphi,\tau)$,
$\hat \lambda(\lambda,\varphi,\tau)$:
\begin{eqnarray*}
\hat I[\hat \varphi(\varphi,\tau),\tau]=I[\varphi,\tau],\quad
(\hat \delta_{\hat \lambda} \hat \varphi)(\varphi,\lambda,\tau)
\approx
\delta_\lambda \hat \varphi(\varphi,\tau),
\end{eqnarray*}
where $\approx$ is `equality on-shell' 
(equality for all solutions to the field equations).

Accordingly, a deformation is called trivial if the deformed action and
gauge transformations are
equivalent to the original action and gauge transformations, i.e., if
$I\sim I^{(0)}$ and 
$\delta\sim \delta^{(0)}$.

We may distinguish
two types of nontrivial deformations:\\[6pt]
{Type I:} $I\not\sim I^{(0)}$, $\delta_{\hat \lambda}\sim\delta_\lambda^{(0)}$, i.e.,
the deformation of the action is nontrivial whereas
the deformation of the gauge transformations is trivial.
\\[6pt]
{Type II:} $I\not\sim I^{(0)}$, $\delta_{\hat \lambda}\not\sim\delta_\lambda^{(0)}$, i.e.,
the deformations of both the action and the gauge transformations are nontrivial.
\\[6pt]
Notice that in this terminology noncommutative Yang-Mills
theories as described in sect. \ref{rev} are type I deformations
of Yang-Mills theories because SW maps are field redefinitions that
bring the noncommutative gauge transformations back to the standard
(commutative) form, i.e., the deformation of the gauge transformations
is trivial.


\subsection{BRST-cohomological approach to
consistent deformations}

The BRST-cohomological approach to
consistent deformations \cite{Barnich:vg}
is most conveniently formulated in
the so-called field-antifield formalism \cite{Batalin:jr}.
The `fields' $\phi^a$ of that formalism 
are the fields $\varphi^i$ occurring in the action $I[\varphi]$,
ghost fields $C^\alpha$ corresponding to the nontrivial gauge
symmetries of the action, as well as 
ghost fields of higher order (`ghosts for ghosts')
if the gauge transformations are reducible.
Each field is accompanied by an antifield $\phi^*_a$ according to
definite rules which are not reviewed here. In particular this
allows one to define the so-called antibracket $(\ ,\ )$ of functions or
functionals of the fields and antifields according to
\begin{eqnarray*}
(F,G)=\int d^nx\, F \Big(
\frac{\stackrel{\leftarrow}{\delta}}{\delta\phi^a(x)}\
\frac{\stackrel{\rightarrow}{\delta}}{\delta\phi^*_a(x)}
-
\frac{\stackrel{\leftarrow}{\delta}}{\delta\phi^*_a(x)}\
\frac{\stackrel{\rightarrow}{\delta}}{\delta\phi^a(x)}
\Big) G.
\end{eqnarray*}
A central object of the formalism is the 
master action $S$. Its importance originates from the fact
that it contains both the action
$I[\varphi]$ and all information about its gauge symmetries,
such as the gauge transformations, their commutator algebra,
reducibility relations etc.
In particular the gauge transformations occur in $S$ via terms
$\varphi^*_i\delta_C\varphi^i$  where $\delta_C\varphi^i$ is
a gauge transformation of $\varphi^i$ with ghost fields $C$ in place of 
gauge parameters $\lambda$. The information about the
gauge symmetry is encoded in the master equation $(S,S)=0$,
\begin{eqnarray*}
S[\phi,\phi^*]=I[\varphi]+
\int d^nx\, \varphi^*_i\delta_C\varphi^i
+\dots\ \mbox{such that}\
\underbrace{\phantom{\int\hspace{-1.5ex}} {(S,S)=0.}}_{{\mathrm{master}}\atop{\mathrm{equation}}}
\end{eqnarray*}
In particular $S$ defines the
BRST differential $s$ via the antibracket with $S$. The master equation
$(S,S)=0$ implies that $s$ squares to zero ($s^2=0$),
\[
s=(S,\ \cdot\ )\quad (\Rightarrow\ s^2=0).
\]
These properties of $S$ make it so useful in the context of consistent
deformations. Indeed, the fact that $S$ contains both the action and the 
gauge transformations
allows one to analyse consistent deformations in terms of the single
object $S$ that has to satisfy the master equation,
\[
S=S^{(0)}+\sum_{k\geq 1}\tau^k S^{(k)},\quad (S,S)=0.
\]
The first relation to BRST-cohomology can be established by differentiation
of the master equation with respect to the deformation parameter:
\[
(S,S)=0\quad 
\stackrel{\frac{\partial}{\partial\tau}}{\Rightarrow}\quad
\Big(S,\frac {\partial S}{\partial\tau}\Big)=0\quad \Leftrightarrow\quad
{s\,\frac {\partial S}{\partial\tau}=0}.
\]
This shows that $\partial S/\partial\tau$ is a cocycle of $s$.
The second relation to BRST-cohomology derives from the
fact that field redefinitions (of $\varphi$ and/or
the gauge parameters) translate into
anticanonical transformations
$\hat \phi(\phi,\phi^*,\tau)$,
$\hat \phi^*(\phi,\phi^*,\tau)$ (these are transformations
generated via the antibracket by some functional $\Xi$).
This implies:
\[
\frac{d\hat \phi}{d\tau}=(\Xi,\hat \phi),\
\frac{d\hat \phi^*}{d\tau}=(\Xi,\hat \phi^*)\ \Rightarrow\
\frac{dS(\hat \phi,\hat \phi^*,\tau)}{d\tau}
=\frac{\partial S}{\partial\tau}-(S,\Xi)
=\frac{\partial S}{\partial\tau}-s\,\Xi.
\]
As a consequence, 
master actions of equivalent deformations
are related as follows:
\[
S\sim S'\quad\Rightarrow\quad
\frac{\partial S}{\partial\tau}-\frac{\partial S'}{\partial\tau}=s\,\Xi.
\]
This shows that 
consistent deformations are determined by the BRST-co\-ho\-mo\-logy $H(s)$
in ghost number 0 since ${\partial S}/{\partial\tau}$ (i) has to be a
BRST-cocycle, (ii) is defined only up to a BRST-coboundary, and (iii) has
ghost number 0 ($S$ has ghost number 0 according to the standard
ghost number assignments).


\subsection{BRST-cohomological analysis of SW maps}

To describe SW maps in the field-antifield formalism we denote the
`noncommutative fields' by $\hat \phi$ and the `commutative' fields by $\phi$.
Actually we enlarge the setup here as compared to
sect. \ref{rev}: all the fields $\hat \phi$ and $\phi$ take values
in the enveloping
algebra of the Lie algebra of the gauge group, resp. some representation $\{T_A\}$
thereof. The superfluous fields $\phi$ (those that do not belong to
the Lie algebra of the gauge group) are set to zero at the end
of the construction, see \cite{Barnich:2002pb} for details.
Dropping again the fermions, we have
\[
\{\hat \phi^a\}=\{\hat A_\mu^A,\hat C^A\},\quad \{\phi^a\}=\{A_\mu^A,C^A\}.
\]
The `noncommutative'
master action reads
\begin{eqnarray*}
S[\hat \phi,\hat \phi^*,\tau]=
\int d^nx\Big[ -\frac 14\,\mathrm{Tr}\, (\hat F_{\mu\nu}\star\hat F^{\mu\nu})
+\hat A^{*\mu}_A\star (\hat D_\mu\hat C)^A
+\hat C^*_A\star(\hat C\star\hat C)^A\Big].
\end{eqnarray*}
The existence of a
SW map means that the gauge transformations can be brought to the
standard Yang-Mills form, which particularly does not depend on $\tau$.
In terms of the master action this means that there is an
anticanonical transformation $\hat \phi(\phi,\phi^*,\tau)$,
$\hat \phi^*(\phi,\phi^*,\tau)$ which casts $S[\hat \phi,\hat \phi^*,\tau]$
in the form of an effective type Yang-Mills action
$I_\mathrm{eff}[A,\tau]$ plus a piece that involves the antifields and encodes gauge transformations
of Yang-Mills type (for the enveloping algebra),
\begin{eqnarray*}
S[\hat \phi(\phi,\phi^*,\tau),\hat \phi^*(\phi,\phi^*,\tau),\tau]
=
\underbrace{\phantom{\int\hspace{-1.5ex}} I_\mathrm{eff}[A,\tau]}_{\mathrm{{no\, antifields}}}+
\int d^nx
\Big[ \underbrace{\phantom{\int\hspace{-1.5ex}} A^{*\mu} D_\mu C
+C^* CC}_{{\mathrm{no\ dependence\ on\ }\tau}}\Big],
\end{eqnarray*}
where indices have been dropped ($A^{*\mu} D_\mu C$ means $A^{*\mu}_A (D_\mu C)^A$ etc).
Differentiating with respect to $\tau$ und
using the properties of anticanonical transformations (see above), we obtain 
\begin{eqnarray*}
\frac{\partial S}{\partial\tau}-s\,\Xi=\frac{\partial I_\mathrm{eff}[A,\tau]}{\partial\tau},
\quad
\frac{d\hat \phi}{d\tau}=(\Xi,\hat \phi),\quad
\frac{d\hat \phi^*}{d\tau}=(\Xi,\hat \phi^*).
\end{eqnarray*}
Hence, in order to find and analyse SW maps
one may analyse whether ${\partial S}/{\partial\tau}$ 
can be written as a BRST-variation $s\Xi$ up to
terms that do not involve antifields. Notice that $\Xi$
gives the SW map. For ${\partial S}/{\partial\tau}$ one obtains
\begin{eqnarray*}
\frac{\partial S}{\partial\tau}=\frac{{\mathrm{i}} \theta^{\alpha\beta}}{2}
\int d^nx\Big[ &&\mathrm{Tr}\, (
-\hat F^{\mu\nu}\star\partial_\alpha \hat A_\mu\star \partial_\beta \hat A_\nu)
\\
&&+\hat A^{*\mu}\star 
\{\partial_\alpha \hat A_\mu\stackrel{\star}{,}\partial_\beta\hat C\}
+\hat C^*\star\partial_\alpha\hat C\star\partial_\beta\hat C\Big],
\end{eqnarray*}
where $\{\ \stackrel{\star}{,}\ \}$ denotes the $\star$-anticommutator,
\[
\{X \stackrel{\star}{,}Y \}=X\star Y+Y\star X.
\] 
This expression for ${\partial S}/{\partial\tau}$ is indeed
BRST-exact up to terms that do not contain antifields. One can infer this
by means of so-called contracting homotopies for derivatives of the ghost fields used already
in \cite{Brandt:1989gy,Dubois-Violette:1992ye}. 
We shall not review the construction of these homotopies here since this is a somewhat technical matter. Rather, we shall only present the result. It is actually ambiguous
as we shall discuss below. In particular it depends on the specific contracting homotopy
one uses (there are various options).
A particularly nice version of the result is
\begin{eqnarray*}
&&\Xi=\frac{{\mathrm{i}}}4\,\theta^{\alpha\beta}\int d^nx\,(
-\hat A^{* \mu}
\{\hat F_{\alpha\mu}+\partial_\alpha \hat A_\mu\stackrel{\star}{,}\hat A_\beta\}
+\hat C^{*}
\{\hat A_\alpha\stackrel{\star}{,}\partial_\beta \hat C\}),
\\
&&
\frac{d\hat A_\mu}{d\tau}=(\Xi,\hat A_\mu)
=\frac{{\mathrm{i}}}4\,\theta^{\alpha\beta}
\{\hat F_{\alpha\mu}+\partial_\alpha \hat A_\mu\stackrel{\star}{,}\hat A_\beta\},
\\
&&
\frac{d\hat C}{d\tau}=(\Xi,\hat C)
=-\frac{{\mathrm{i}}}4\,\theta^{\alpha\beta}
\{\hat A_\alpha\stackrel{\star}{,}\partial_\beta \hat C\}
\\
&&\frac{d I_\mathrm{eff}[\hat A(A,\tau),\tau]}{d\tau}=
{\mathrm{i}}\theta^{\alpha\beta}\!\int d^nx\,\mathrm{Tr}\,
(\frac 18 \hat F_{\alpha\beta}\star \hat F_{\mu\nu}\star \hat F^{\mu\nu}\!
-\frac 12\hat F_{\alpha\mu}\star \hat F_{\beta\nu}\star \hat F^{\mu\nu}).
\end{eqnarray*}
The expressions for ${d\hat A_\mu}/{d\tau}$ and ${d\hat C}/{d\tau}$ are differential equations for SW maps
of the same form as derived in \cite{Seiberg:1999vs} for $U(N)$-theories.

The
ambiguities of the result can be described in terms of $\Xi$ as
shifts $\Xi+\Delta\Xi$ of $\Xi$ which satisfy
\begin{eqnarray*}
0=s\,(\Delta\Xi) + \mbox{terms without antifields,}
\end{eqnarray*}
where the terms without antifields yield the shift $d(\Delta  I_\mathrm{eff})/d\tau$
corresponding to $\Delta\Xi$.
This is again an equation that can be analysed by cohomological means
which are not reviewed here, and we 
only present the result:
the general SW map $\hat A_\mu(A,\tau)$, $\hat \lambda(\lambda,A,\tau)$
for the gauge fields and gauge parameters can be written as
\begin{eqnarray*}
\hat A_\mu(A,\tau) &=& \Big[\Lambda^{-1}\star\hat A_\mu^{sp}\star\Lambda
          +\Lambda^{-1}\star\partial_\mu\,\Lambda\Big]_{A_\mu\rightarrow 
          A'_\mu(A,\tau)}
\\
\hat \lambda(\lambda,A,\tau)&=&  
          \Big[\Lambda^{-1}\star\hat \lambda^{sp}\star\Lambda
          +\Lambda^{-1}\star\delta_\lambda\Lambda\Big]_{A_\mu\rightarrow 
          A'_\mu(A,\tau)}
\end{eqnarray*}
where
\begin{eqnarray*}
&&\Lambda(A,\tau)=\exp_\star(f^B(A,\tau)T_B)\ \mbox{with arbitrary}\  
f^B(A,\tau),
\\[8pt]
&&\hat A_\mu^{sp}(A,\tau),\ \hat \lambda^{sp}(\lambda,A,\tau)\
\mbox{is a particular SW map,}
\\[8pt]
&&A_\mu^{\prime\, B}(A,\tau)=[A_\mu+W_\mu(A,\tau)]^C R_C^B(\tau)\ \mbox{where:}
\\[8pt]
&&\delta_\lambda W_\mu(A,\tau)=[W_\mu(A,\tau),\lambda]\ 
\mbox{(i.e., $W_\mu$ is gauge covariant),}
\\[8pt]
&&T_B\rightarrow R_B^C(\tau)\, T_C\ \mbox{is an (outer) Lie algebra automorphism}.
\end{eqnarray*}
Recall that $\{T_A\}$ is (a representation of) the enveloping algebra of the Lie algebra of the gauge group.
Hence, the Lie algebra automorphisms $T_B\rightarrow R_B^C(\tau) T_C$ that enter here refer to the
Lie algebra of $\{T_A\}$ rather than to the Lie algebra of the gauge group.
Without loss of generality one may restrict these automorphisms to
outer automorphisms since inner ones are already covered by the $\Lambda$-terms.
Note that the latter are (field dependent) noncommutative gauge transformations of 
a special SW map $\hat A_\mu^{sp}$.

Hence, SW maps are determined only up to (compositions of)
noncommutative gauge transformations of $\hat A_\mu$,
gauge covariant shifts of enveloping algebra valued gauge fields $A_\mu$, and 
outer automorphisms of the Lie algebra of the enveloping algebra.

\section{Gauge anomalies}\label{sect4}

A 1-loop computation, performed with dimensional
regularization, yields the following
expression for gauge anomalies
in four-dimensional noncommutative Yang-Mills theories with chiral fermions \cite{Brandt:2003fx}:
\beq
{\cal A} [\hat C,\hat A,\tau]=\int \mathrm{Tr}[\hat C\star d (\hat A\star d\hat A+\frac 12 \hat A\star\hat A\star\hat A)],
\label{a1}\eeq
where we used differential form notation ($d=dx^\mu\partial_\mu$, $\hat A=dx^\mu\hat A_\mu$).
This expression is reminiscent of anomalies in ordinary (commutative) Yang-Mills theories since it arises from the latter by replacing commutative fields $C$ and $A_\mu$ with their noncommutative counterparts and ordinary products with $\star$-products. However, the presence of $\star$-products poses an apparent puzzle:
${\cal A}=0$ does not only impose the usual anomaly cancellation conditions $\mathrm{Tr}(T_{(a} T_{\phantom{)}\!\!b} T_{c) })=0$ but additional conditions at higher orders in $\theta$, such as $\mathrm{Tr}(T_{[a} T_{\phantom{]}\!\!b} T_{c] })=0$. On the other hand
all candidate gauge anomalies of noncommutative Yang-Mills theories of the type considered here are known because these theories can be considered 
Yang-Mills theories of the effective type whose anomalies were exhaustively classified (see \cite{Barnich:2000zw}
for a review).
These known results state in particular that the chiral (Bardeen) anomalies exhaust all candidate gauge anomalies when the gauge group is semisimple. According to this result (\ref{a1}) is cohomologically equivalent to a standard chiral anomaly, i.e., all (infinitely many!) $\theta$-dependent terms in (\ref{a1}) are BRST-exact when the gauge group is semisimple. 

The situation is more involved when the gauge group contains an abelian factor. In this case there are additional, and in fact infinitely many, candidate anomalies, and it is not obvious from the outset whether or not some of them occur in (\ref{a1}). The question is thus: is (\ref{a1}) always cohomologically equivalent to a standard chiral anomaly, even when the gauge group contains abelian factors? The answer is affirmative, as was shown in \cite{Brandt:2003fx}.
Again, we shall only briefly sketch how this result was obtained and drop all details.

The idea is to differentiate (\ref{a1}) with respect to $\tau$ and to show that the resultant expression is BRST-exact. The reason for dealing with
${d{\cal A}}/{d\tau}$ rather than with ${\cal A}$ itself is that, as it turns out, ${d{\cal A}}/{d\tau}$ is the BRST-variation of an
expression that can be compactly written as an integrated $\star$-polynomial of the noncommutative variables
$\hat A_\mu$:
\begin{eqnarray*}
\frac{d{\cal A}}{d\tau}&=&s{\cal B}_\star\, ,\\
{\cal B}_\star&=&\frac{{\mathrm{i}}\theta^{\alpha\beta}}{2}
\int\,\mathrm{Tr}\,( \hat A_\alpha\star\partial_\beta d\hat A\star d\hat A
-\frac 12\, d\hat A_\alpha\star \hat A_\beta\star d\hat A\star \hat A
\\
&&
+\frac 32\, d\hat A\star d\hat A_\alpha\star\hat A\star\hat A_\beta
-\frac 12\, d\hat A_\alpha\star \hat A_\beta\star \hat A\star  d\hat A
\\
&&
+\partial_\alpha\hat A_\beta\star d\hat A\star \hat A\star \hat A
+ \mbox{terms with 5 or 6 $\hat A$'s}).
\end{eqnarray*}
We remark that ${\cal B}_\star$ is not unique (it is determined only up to BRST-cocycles with ghost number 0) and
can be written in various ways. Hence, the expression given above is just one particular choice.
The desired result for ${\cal A}$ is now obtained using ${\cal A}(\tau)={\cal A}(0)+\int_0^{\tau} d\tau' {d{\cal A}}/{d\tau'}$. 
This gives
\beq
{
{\cal A}
=\int \mathrm{Tr}[C d (A dA+\frac 12 A^3)]
+s\,{\cal B}[A,\tau],\quad
{\cal B}[A,\tau]=\int_0^\tau d\tau'\ {\cal B}_\star[\hat A(A,\tau'),\tau'].
}
\label{a2}\eeq
Notice that ${\cal B}$, in contrast to ${\cal B}_\star$, can not be naturally written 
as an integrated $\star$-polynomial of the noncommutative variables
$\hat A_\mu$ because of the dependence of $\hat A(A,\tau')$ on $\tau'$.
(\ref{a2}) shows that ${\cal A}$ is indeed given by the standard chiral gauge anomaly
$\int \mathrm{Tr}[C d (A dA+\frac 12 A^3)]$ up to a BRST-exact piece $s{\cal B}$. 
Hence, at least at the 1-loop level, noncommutative Yang-Mills theories do not
possess additional gauge anomalies or anomaly cancellation conditions as compared to
the corresponding commutative theories, even when the gauge group contains abelian factors
(the above results apply to all gauge groups). Notice that
$-{\cal B}$ is the counterterm that cancels the $\theta$-dependent terms in ${\cal A}$.

%


\begin{thebibliography}{99.}
%
%
%


\bibitem{Jurco:2000ja}
B.~Jurco, S.~Schraml, P.~Schupp and J.~Wess,
Eur.\ Phys.\ J.\ C {\bf 17}, 521 (2000)
[arXiv:hep-th/0006246].

\bibitem{Seiberg:1999vs}
N.~Seiberg and E.~Witten, 
JHEP {\bf 09}, 032 (1999)
[arXiv:hep-th/9908142].

\bibitem{Barnich:2002tz}
G.~Barnich, F.~Brandt and M.~Grigoriev,
Fortsch.\ Phys.\  {\bf 50}, 825 (2002)
[arXiv:hep-th/0201139].

\bibitem{Barnich:2002pb}
G.~Barnich, F.~Brandt and M.~Grigoriev,
JHEP {\bf 08}, 023 (2002)
[arXiv:hep-th/0206003].

\bibitem{Barnich:2003wq}
G.~Barnich, F.~Brandt and M.~Grigoriev,
Nucl.\ Phys.\ B {\bf 677}, 503 (2004)
[arXiv:hep-th/0308092].

\bibitem{Brandt:2003fx}
F.~Brandt, C.~P.~Mart\'{\i}n and F.~Ruiz Ruiz,
JHEP {\bf 07}, 068 (2003)
[arXiv:hep-th/0307292].

\bibitem{Barnich:2001mc}
G.~Barnich, M.~Grigoriev, and M.~Henneaux, 
JHEP {\bf 10}, 004 (2001)
[arXiv:hep-th/0106188].

\bibitem{Barnich:vg}
G.~Barnich and M.~Henneaux,
Phys.\ Lett.\ B {\bf 311}, 123 (1993)
[arXiv:hep-th/9304057].

\bibitem{Batalin:jr}
I.~A.~Batalin and G.~A.~Vilkovisky,
Phys.\ Lett.\ B {\bf 102}, 27 (1981).

\bibitem{Brandt:1989gy}
F.~Brandt, N.~Dragon and M.~Kreuzer,
Nucl.\ Phys.\ B {\bf 332}, 224 (1990).

\bibitem{Dubois-Violette:1992ye}
M.~Dubois-Violette, M.~Henneaux, M.~Talon and C.~M.~Viallet,
Phys.\ Lett.\ B {\bf 289}, 361 (1992)
[arXiv:hep-th/9206106].

\bibitem{Barnich:2000zw}
G.~Barnich, F.~Brandt and M.~Henneaux,
Phys.\ Rept.\  {\bf 338}, 439 (2000)
[arXiv:hep-th/0002245].

\end{thebibliography}
\end{document}